\documentclass[]{spie}  %>>> use for US letter paper
\usepackage{amsmath,amsfonts,amssymb}
\usepackage{graphicx}
\usepackage[colorlinks=true, allcolors=blue]{hyperref}

\title{Pupil plane WFSs for LGS systems of giant telescopes: the case of Ingot}

\author[a,d]{Elisa Portaluri}
\author[b,c,d]{Roberto Ragazzoni}
\author[c,d]{Davide Greggio}
\author[c,d]{Carmelo Arcidiacono}
\author[c,d]{Maria Bergomi}
\author[c,d]{Simone Di Filippo}
\author[c,d]{Marco Dima}
\author[c,d]{Jacopo Farinato}
\author[b,c,d]{T\^ania Sofia Gomes Machado}
\author[c,d]{Demetrio Magrin}
\author[c,d]{Kalyan Kumar Radhakrishnan Santhakumari}
\author[c,d]{Valentina Viotto}

\affil[a]{INAF - Osservatorio Astronomico d'Abruzzo, via Mentore Maggini snc, I-64100 Teramo, Italy}
\affil[b]{Dipartimento di Fisica e Astronomia ``G. Galilei'', Universit\`a di Padova, vicolo dell'Osservatorio 3, I-35122 Padova, Italy }
\affil[c]{INAF - Osservatorio Astronomico di Padova, vicolo dell'Osservatorio 5, I-35122 Padova, Italy}
\affil[d]{ADONI - Laboratorio Nazionale Ottica Adattiva (Italy)}

\authorinfo{Further author information: (Send correspondence to E.P.)\\E.P.: E-mail: elisa.portaluri@inaf.it}

\begin{document} 
\maketitle

\begin{abstract}
The Ingot WFS belongs to a class of pupil-plane WFSs designed to address the challenges posed by Sodium Laser Guide Stars, and consists of a combination of refractive and reflective surfaces, arranged into a complex prismatic shape that extends in three dimensions. Specifically, it leverages the Scheimpflug principle to sense the full 3D volume of such elongated, time-varying sources, thus optimizing the performance of the next-generation AO-assisted giant telescopes. In this work we discuss the geometrical and optical motivations endorsing the development of this class of WFSs, showing the different configurations we propose to the AO community. We also provide a first order comparative analysis with other approaches and review the state-of-the-art of the Ingot project, including improvements made in the laboratory and future milestones.  
\end{abstract}

\keywords{Astronomical instrumentation, methods and techniques 
-  Instrumentation: adaptive optics 
 - Instrumentation: detectors
 - LGS - Ingot WFS - LGS elongation -  LGS truncation - ELT}

\section{INTRODUCTION}
\label{sec:intro}  % \label{} allows reference to this section

From the beginning, the telescopes enabled the humanity to push the limits of seeing the Universe with naked eyes, changing the perception of our place. From then, we always wanted to go further, facing challenges and building even more advanced and sophisticated instrumentation.
Thanks to HST before and JWST now, we are experiencing a succession of groundbreaking scientific discoveries, going from the better understanding of the already known to enabling newer explorations of the cosmos and unknown.  
But we want and we can go deeper and sharper than now: the combination of a larger aperture with the Adaptive Optics (AO) systems will provide unparalleled data, making the next-generation class of telescopes the most powerful and coveted instruments in the JWST era.
This is the intention of the European Extremely Large Telescope (ELT\cite{Gilmozzi2007}), the Giant Magellan Telescope (GMT \cite{Johns2008}), and the Thirty Meter Telescope (TMT\cite{Szeto2008}): to see what the others cannot. 

In this context, the technology advancements should follow a steep ascent leading to the future needs of the systems, not being just barely adapted to the case, as this lack of perspective will make the expected performance suboptimal. For example, a proposed option for enlarging the sky coverage without the use of Laser Guide Stars (LGSs) is the Global-MCAO approach\cite{Portaluri2017, Portaluri2020}, which would avoid the use of a sensor not really conceived for such a kind of sources.
On the opposite side, we can focus on a possible alternative to the planned Shack-Hartmann wavefront sensor (WFS): in this sense, the development testing, and on-sky verification of a new class called Ingot is absolutely worthy and timely.

In brief, we aim at the complete characterization and verification on-sky of the Ingot WFS, which is designed to cope with the adopted configuration of the ELT AO systems, overcoming the limits of conventional WFSs and, thus, allowing to fully exploit their extraordinary capabilities. 

The geometrical description of this class of WFS, with the recipe to design it for a given generic system in the most reasonably complex version and in the simplest one, is detailed in Ragazzoni et al. (2024)\cite{Ragazzoni2024}, while here we give a resume of the concept (Section~\ref{sec:ingot}).
Numerical simulations and optical bench tests are being developed and ongoing, and are described in Section~\ref{sec:project}, where an overview of the project is given.

\section {RATIONALE OF THE PROJECT} 
\label{sec:ingot}
The departure of the appearance of LGSs from the Natural Guide Stars (NGSs) is somehow proportional to the diameter of the observing telescope, so that this issue has a minimal impact on current 8 to 10 m state-of-the-art class telescopes, but it is expected to play a more significant role in the future gigantic ground based telescopes.
In fact, treating the LGSs as a cylindrical emitting volume rather than point sources reveals several sub-optimal characteristics in the associated WFSs, leading to an increase in the required returning flux to attain a certain level of accuracy.
Most of the literature dealt with such an elongation, managing a number of countermeasures or specific technical issues, such as the truncation of its image when treated by a classical WFS \cite{Vieira2018, Clare2020}.
However, in principle, the conceptual design of a WFS could incorporate the way this light is produced, tailoring that specific need, and thus being more efficient than treating LGSs as stars\cite{Ragazzoni2017}.

The ingot WFS is an innovative and indispensable pupil plane wavefront sensor, proposed by Ragazzoni et al. (2017)\cite{Ragazzoni2017} and refined by the AO Group @INAF-Padova, to be used in combination with LGSs.
The optimal use of the detector is one of the advantages that is fully retained with a pupil plane WFS, and, above all, in principle, this is totally independent of the diameter of the telescope aperture.
This makes the Ingot WFS attractive for ELTs, especially if the lack of large format detectors would lead to the need for truncation of the light, as occurs with the traditional Shack-Hartmann (SH) WFSs. Therefore, by extending the pyramid WFS concept\cite{Ragazzoni1996} in order to cope with the typical elongation of LGSs, the performance of the AO system would significantly benefit of it (Fig.1).
\begin{figure} [ht]
   \begin{center}
   \begin{tabular}{c} 
   \includegraphics[height=9cm]{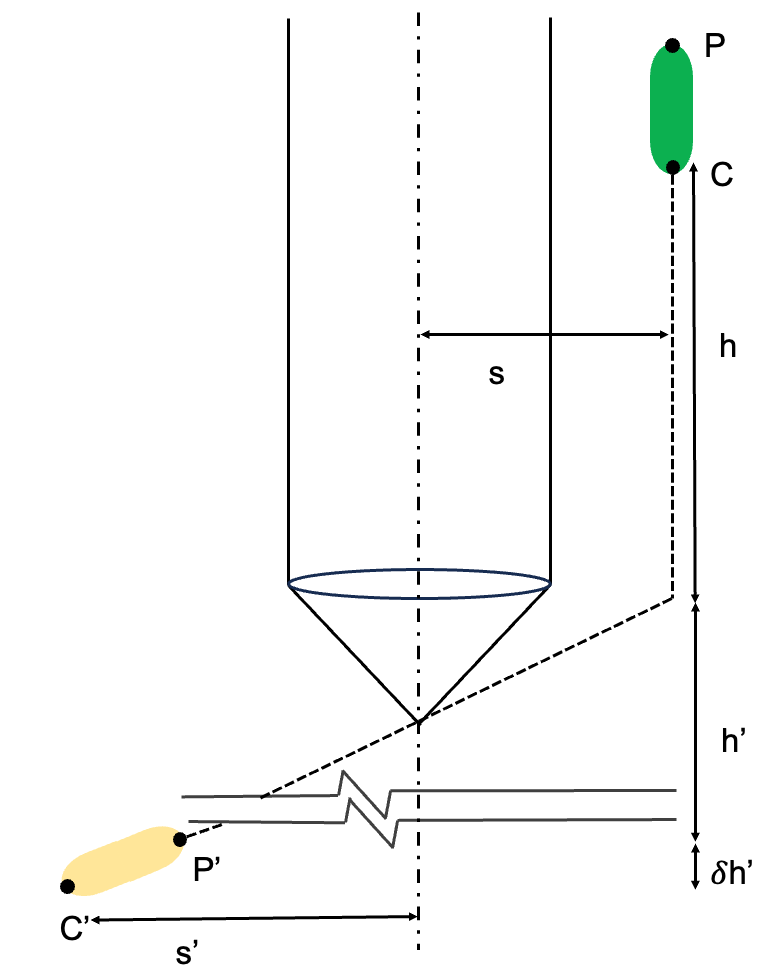}
   \end{tabular}
   \end{center}
   \caption[example] 
   { \label{fig:ELT} 
Example of an ELT system working with a LGS fired at an height $h$ by a launched at a distance $s$ from the centre of the telescope. The points $P$ and $C$, which indicates the upper and lower limits of the light, are reimaged into $P'$ and $C'$ on a volume at a distance $h'$ ($+ \delta h'$) and $s'$.}
   \end{figure}

 The Ingot WFS is based on the idea of optimizing the match between the elongated source, which is fired from the side of the primary mirror and can be considered like “a cigar in the sky”, and the optical surface used to sense it, shaped like an ingot prism (Figure~\ref{fig:ELT}). In fact, it consists of a combination of refractive and reflective surfaces arranged onto a complex prismatic shape that extends in three dimensions.  
We highlight that all the geometrical demonstrations made in our series of papers are in the case of a laser launcher outside the telescope, which is the configuration studied for the Ingot WFS. In that case, the light fired will never enter the image of the LGS itself, while if the launcher is behind the primary mirror, that light is incorporated into the image.
 
\begin{figure} [ht]
   \begin{center}
   \begin{tabular}{c} 
   \includegraphics[height=5cm]{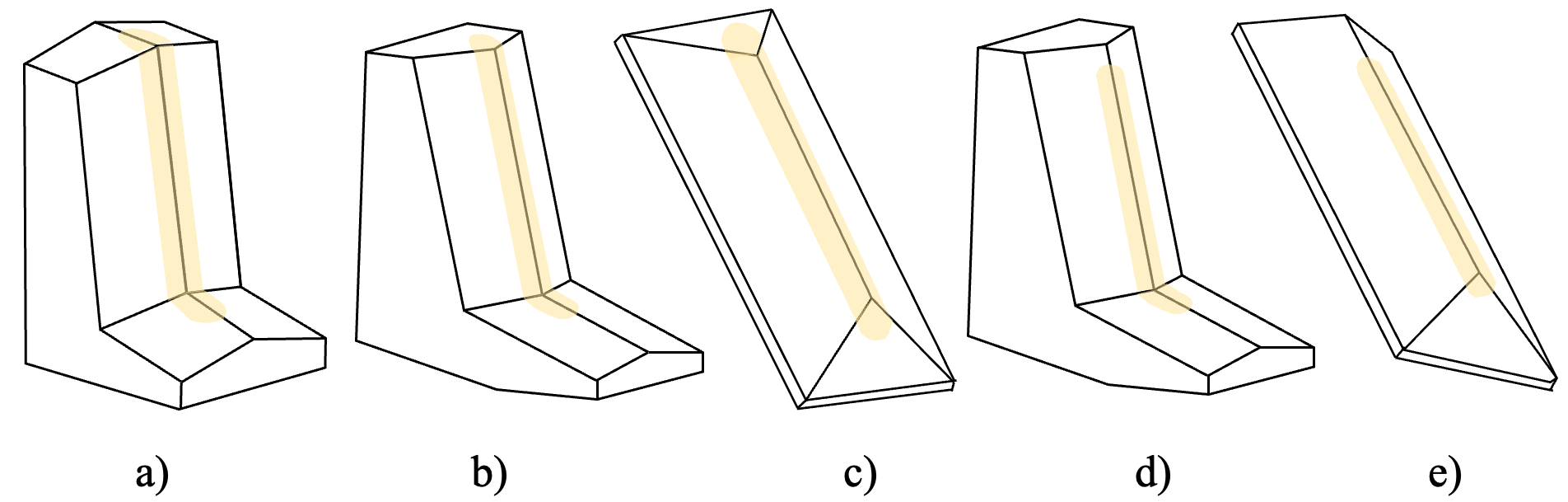}
   \end{tabular}
   \end{center}
   \caption[example] 
   { \label{fig:prisms} 
Different optomechanical layouts of the ingot prisms with 6 (a), 5 (b), 4 (c and d) and 3 (e) pupils. An example of the LGS light intersecting the prism on the different (refractive and/ord reflective) faces is also shown.}
   \end{figure}

 The ingot prism is the optical perturbator to be introduced at the reimaged volume, where an LGS is being focused by a large telescope and, after that, a common collimator would produce a number of pupil images that can be used to estimate the derivatives of the WF.
Several optomechanical layouts of the ingot are possible, going in a range from 3 to 6 reimaged pupils, according to the desirable rationale and accounting for an almost negligible deterioration of the performance, especially if compared with other WFSs used with LGSs (Figure~\ref{fig:prisms}).
Some of the possible implementations and variations among the class are particularly easy to realize, and in this project, we focus on the configuration with 3 pupils (case e).

 However, only the light coming from the edges of the LGS is used to sense the derivative in the direction perpendicularly to the elongation, while the light coming from the entire LGS is used to measure the derivative in the direction of the elongation. This information is somehow computed from the 3 to 6 pupils, combining their intensities in some way described in \citenum{Portaluri2019}. One can argue that in the 3 pupils version the error associated with the y-derivative of the wavefront will be much bigger, having the same limitation of a SH WFS using a smart centroid algorithm to limit the noise the effects on centroid measurements and a enough big detector.
 However, this issue can be resolved by combining the lights from different LGSs, as a system like the ELT one will account for 6 LGSs. 
 Moreover, the lower limit of the Sodium layer is  sharper and brighter than the upper one, which is close to the sodium distribution tale (see some examples in Ref.~\citenum{GomesMachado2024}): the associated amount of light is probably very low, so the upper limit would probably not bring much more information in y-direction derivative computation. That is another point going into the direction of using a version with only 3 pupils to make the optics and mechanics easier. In fact, as a first interaction with the companies for building the ingot prism, we were directed in gluing together several pieces, and in this sense, less complexity means less components to be put one each
other. 

\begin{figure} [ht]
   \begin{center}
   \begin{tabular}{c} 
   \includegraphics[height=9cm]{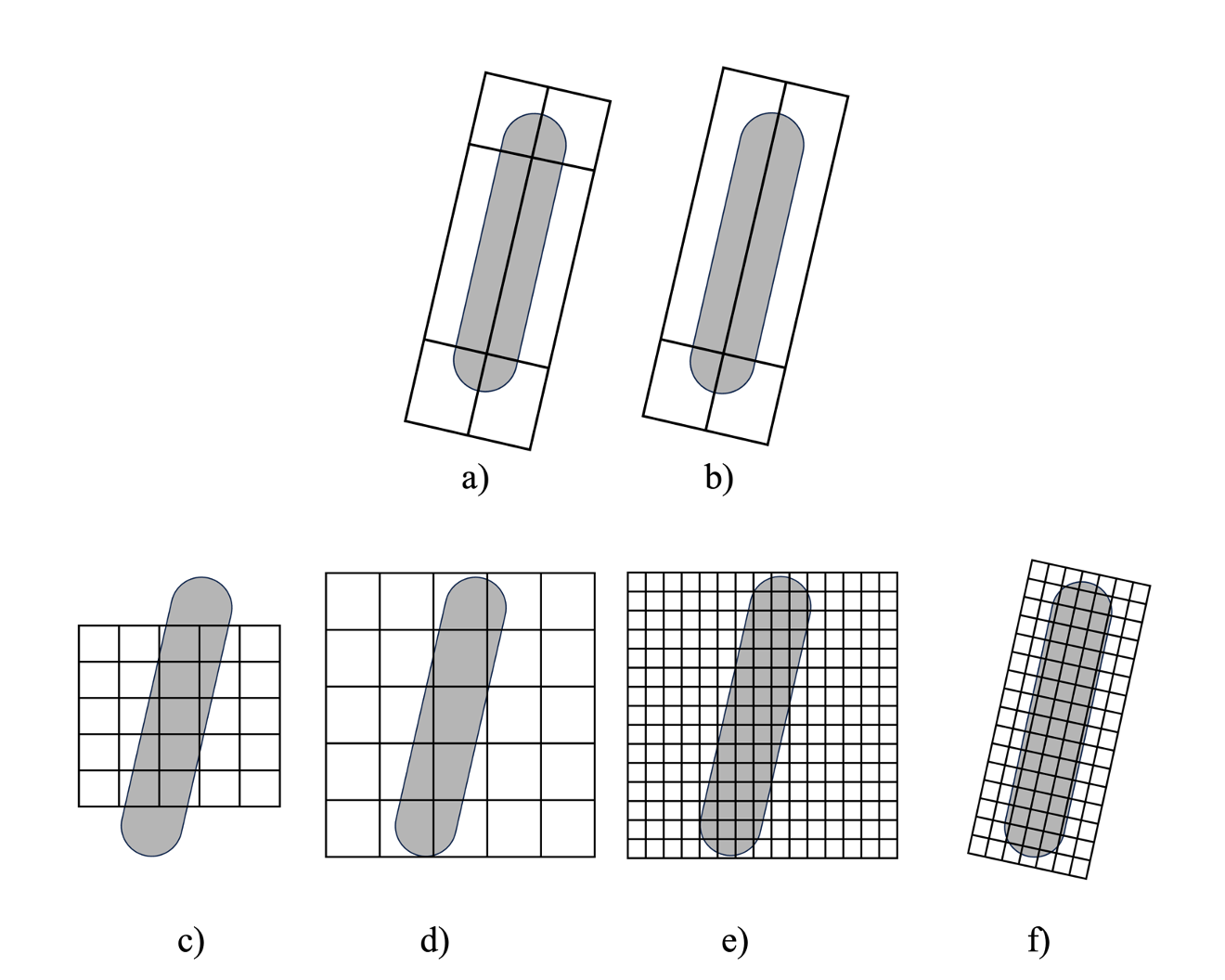}
   \end{tabular}
   \end{center}
   \caption[example] 
   { \label{fig:WFScomparison} 
Top panel: Elongated spot on the ingot (pupil plane) WFS for the 6 (a) and 3 (b) pupils versions. Bottom panel: images on the focal plane for SH WFS of an elongated spot produced by a subaperture far away from the location of the launcher. The 4 different cases are represented: the need of truncation because the spot is bigger than the detector (c); the need of a very big detector having the same number of pixels of the previous one (d) and with a finer sampling (e) and the use of a polar detector to optimize the geometrical configuration (f). }
   \end{figure}

Finally, Figure~\ref{fig:WFScomparison} reports a comparison with a SH under general conditions, considering characteristic cases that give a hint on how comparable is the sensitivity in the two examples, also recalling the case of the polar CCD. Clearly, this is not a detailed and comprehensive comparison that would depend upon a number of conditions (format and RON of the
CCD, arrangement of the subapertures, size of the aperture and distance of the
projector, etc.) and, of course, we can just mention passing by the ability to use, in a
multiple LGSs, MCAO-like approach, the use of a LGS for the derivative along different
directions. However, this is an illustrative example of the advantages that the ingot WFS will have in the ELT system.
 
\section{PROJECT STATUS}
\label{sec:project}
As described in Ref.~\citenum{Portaluri2022, Portaluri2023}, we started a
feasibility study for the 3-pupils ingot to explore its sensitivity and resolution, assessing its implementation on large telescopes.
A preliminary evaluation of the performance are been carried out\cite{Portaluri2019}, by using numerical simulation tools\cite{Viotto2018}, which make use of a hybrid approach between a pure ray tracing technique and a Fourier analysis\cite{Viotto2019}.

Furthermore, we setup a laboratory bench to test the 3-pupils ingot sensitivity\cite{2020SPIEKalyan} and develop an automatic procedure to align the system (LGS+WFS)\cite{Difilippo2019, DiFilippo2022}.
The simulated source has a brightness distribution along the line of propagation that can be properly adjusted to mimic the actual measurements\cite{GomesMachado2024}.
Simulations of closed-loop procedures on an optical bench have also been carried out with a simplified model of the ingot\cite{Arcidiacono2020}.

As a general remark, while in the lab we only tested/are testing the 3-pupils configuration, the simulation code was written with the possibility of selecting 3 different options\cite{Viotto2018}: 6, 4, and 3 pupils. A proper analysis of the results and a comparison between them is still ongoing and not really included in the discussion of this paper.

\section{CONCLUSIONS}
\label{sec:conc}
Full sky coverage AO requires the adoption of LGSs, as traditional methods rely on bright natural references close to the science target, limiting the investigation to specific (and small) areas.
With ELT large apertures, the apparent elongation of the beacons is absolutely significant. However, WFS are barely adapted and used in suboptimal mode.
We are developing a new class of WFS in order to optimize the correction for the atmospheric turbulence and, therefore, fully exploit the ELTs capabilities.
They are designed to maximize the wavefront sensing efficiency with LGSs: the two main advantages of these WFS with respet to classical suboptimal solutions are the reduced size of the required detector associated with the wavefront sensor in the case of ELT application and the solution of the truncation problem. These two issues, which challenge the classical SH solution adopted up to now in ELT instrumentation, are intimately related to each other.

The exact recipe to design the wavefront sensor for a given generic system is given in Ref.~\citenum{Ragazzoni2024}, and it reports two quantitative and numerical examples of applications to the ELT telescope, considering two of the five ingot versions described also in this paper.

Simulations and Laboratory tests are still on-going and the future plan foresees the use of available facilities to verify the sensor on sky.

\acknowledgments % equivalent to \section*{ACKNOWLEDGMENTS}       
 
 We acknowledge the ADONI Laboratory for the support in the development of this project. 
 Funding to support the lab activities described in this paper came also from the INAF Progetto Premiale "Ottica Adattiva Made in Italy per i grandi telescopi del futuro".
 E.P. acknowledges the Osservatorio Astronomico di Padova for the hospitality while this paper was in progress.

% References
\bibliography{main} % bibliography data in report.bib
\bibliographystyle{spiebib} % makes bibtex use spiebib.bst

\end{document}